\documentclass[dvips]{arxstspdf}
\usepackage{graphicx}
\usepackage{flushend}
\usepackage{stfloats}

\volume{25}
\issue{4}
\pubyear{2010}
\firstpage{450}
\lastpage{457}
\doi{10.1214/08-STS270}

\begin{document}
\begin{frontmatter}

\title{The EM Algorithm in Genetics, Genomics and Public Health}
\runtitle{The EM Algorithm in Genetics, Genomics and Public Health}

\begin{aug}
\author[a]{\fnms{Nan M.} \snm{Laird}\corref{}\ead[label=e1]{laird@hsph.harvard.edu}}
\runauthor{N. M. Laird}

\affiliation{Harvard School of Public Health}

\address[a]{Nan M. Laird is Professor,
Department of Biostatistics, Harvard School of Public Health,
655 Huntington Avenue, Boston, Massachusetts 02115, USA \printead{e1}.}
\end{aug}

%
\begin{abstract}
The popularity of the EM algorithm owes much to the 1977 paper by
Dempster, Laird and Rubin. That paper gave the algorithm its name,
identified the general form and some key properties of the algorithm
and established its broad applicability in scientific research. This
review gives a nontechnical introduction to the algorithm for a
general scientific audience, and presents a few examples
characteristic of its application.
\end{abstract}

%
\begin{keyword}
\kwd{Incomplete data}
\kwd{maximum likelihood}
\kwd{gene counting}
\kwd{linkage analysis}
\kwd{finding regulatory motifs}
\kwd{diffusion batteries}
\kwd{particle size distributions}.
\end{keyword}
\pdfkeywords{Incomplete data,
maximum likelihood,
gene counting,
linkage analysis,
finding regulatory motifs,
diffusion batteries,
particle size distributions}

\end{frontmatter}

\section{Introduction}

Incomplete data arise in many different settings in the empirical
sciences. An obvious example of incomplete data is missing data,
where multiple measurements are made on each subject, but some
subjects are not observed on all measurements. Many applications
are more subtle and consist of problems where the observed data only
shed light on ``hidden'' or ``latent'' traits which are of primary
interest. This frequently occurs in the engineering setting with
reconstruction or indirect measurement, such as medical imaging with
emission and transmission tomography. The example from public
health that is diagrammed in Box 4 is another example of indirect
measurement. Many applications involve locating clusters of
observations with similar features; distinguishing features are
functions of the observed data, but cluster membership must be
inferred. Such problems occur prominently in bioinformatics and
speech recognition. Finally, some statistical models, such as
variance components or random effects, can be reformulated as
missing data problems simply to make computations easier, even
though no data are missing.

Here we discuss the EM algorithm (Dempster, Laird and Rubin, \citeyear{demLaiRub77},
henceforth DLR), which is designed for computations in a broadly
defined incomplete data setting; it is widely used in many different
areas in the empirical sciences. There are numerous applications
(Smith, \citeyear{smi57}; Hasselblad, \citeyear{has66};
Baum et al., \citeyear{baumETAL70};
Orchard and Woodbury, \citeyear{orcWoo72},
to mention a few) of the algorithm that predate the
DLR paper, but that paper described some key features of the
algorithm that underlie its widespread popularity. First was the
recognition of the generality of the algorithm, and a probabilistic
definition of incomplete data that can be applied very broadly in
different settings. Secondly, the paper provided a simple and
intuitive description of the algorithm and named it
``Expectation--Maximization,'' or EM for short, to reflect the two
steps that comprise its essential nature. There are many technical
descriptions of the algorithm and its properties (see, e.g.,
McLachlan and Krishnan, \citeyear{mclKri97}), and a variety of generalizations. The
purpose of this note is to provide an intuitive description of how
the algorithm works and give four short examples from genetics,
genomics and public health.

We first give a heuristic characterization of the algorithm; the
remainder of the introduction discusses its formulation in more
detail. The essential idea of the EM is to postulate the
availability of additional data (e.g., the values of the
missing measurements in the missing data setting) that make the
estimation problem easy. The EM then proceeds by alternating between
two steps: ``fill in the additional data (E-step)'' and ``estimate the
parameters using the filled in data (M-step).'' This two-step
process is repeated until convergence. DLR put the algorithm on a
rigorous foundation by spelling out how the two steps should be
implemented in a general setting and showing that it maximizes an
objective function.

To describe the algorithm, the first task is to identify a ``complete
data'' version of the problem. This is often the most creative part
of the application, since once a complete data analogue has been
identified for the observed data, the application of the EM is
straightforward. There generally will not be a unique representation
for the complete data, but often there is an ``obvious'' one. For
example, in the case of missing data, the complete data is best
described as the observed data, plus the missing observations; in
this way, the complete data specifies a complete set of measurements
for each subject. In other settings, the complete data will consist
of the data that are observed plus some additional information that
would make the problem easy; the bioinformatics example we describe
in Box 3 and clustering examples in general fall into this category.
At one extreme, the complete data is sometimes best described as
just the missing data, because the observed data provides no
additional information when the missing data are available. As an
example, consider our first example (Box 1) of estimating gene
frequencies. It would be straightforward to estimate the
frequencies of different gene variants (termed alleles), if we
directly observed individual genotypes; in the absence of genotype
information, we use data on traits related to the genotype. Here
the complete data are the observed genetic traits \emph{and} the
genotypes of individuals in the sample (Box 1), but if we know the
genotypes, the observed traits contribute no additional information
about gene frequencies.

Having identified the complete data, we can now delineate the E- and
M-steps of the algorithm. Although logically the E-step precedes
the M-step in the computations, hence EM, conceptually it is easier
to define the M-step first.
\begin{longlist}
\item[\textit{M-step}:] The way we define the complete data determines the
M-step of the algorithm. At the M-step, we obtain our estimates of
the parameters of interest assuming we have observed the possibly
hypothetical ``complete data.'' Formally the M-step (for Maximum
likelihood) uses the complete data to obtain maximum likelihood
estimates of the parameters. In many instances, this will be
simple, familiar statistics, that is, means, variances and covariances,
or proportions; each of the four examples we will discuss is based
on a complete data multinomial likelihood and just requires
estimating probabilities from sample frequencies at the M-step.
Exactly how this M-step is carried out depends upon the application,
but it is worth noting that the ease of computations depends in
large measure on defining the complete data so that performing the
M-step is easy.
\item[\textit{E-step}:] Once the M-step is done, we have interim estimates of
the relevant parameters which can now be used, along with the
observed data, to calculate expected values for the ``missing data,''
or technically, computing the expected log-likelihood of the
complete data. Again, the exact nature of the \mbox{E-step} (for computing
the Expected log-likelihood) is application dependent; in each of
the examples we discuss, the \mbox{E-step} involves the computation of
conditional probabilities. These two steps are iterated until
convergence. Although the algorithm is not guaranteed to maximize
the likelihood function, it has some attractive numerical
properties. These include increasing the likelihood at each
iteration and a guarantee that the parameter estimates will remain
in the boundary space, that is, probabilities will always be between 0
and 1, and variance--cova\-riance matrices will be positive semi-definite.
\end{longlist}

\section{Examples from Genetics}

The genetics literature is replete with examples of the EM. Before
genotypes were readily available via modern technologies, the EM was
often used to estimate gene frequencies from data on associated
Mendelian traits. An individual's genotype consists of a pair of
alleles, one inherited from each parent. Even when the genotypes of
individuals are observed, there are still many important estimation
problems with naturally occurring incompleteness, especially if it
is important to determine which allele is inherited from which
parent. One example is the reconstruction of haplotypes, that is,
the set of alleles at different loci all lying on the same
chromosome, from pairs of alleles at the different loci. A second
example that we will discuss is estimating allele sharing in a pair
of affected siblings.

\begin{figure}

\includegraphics{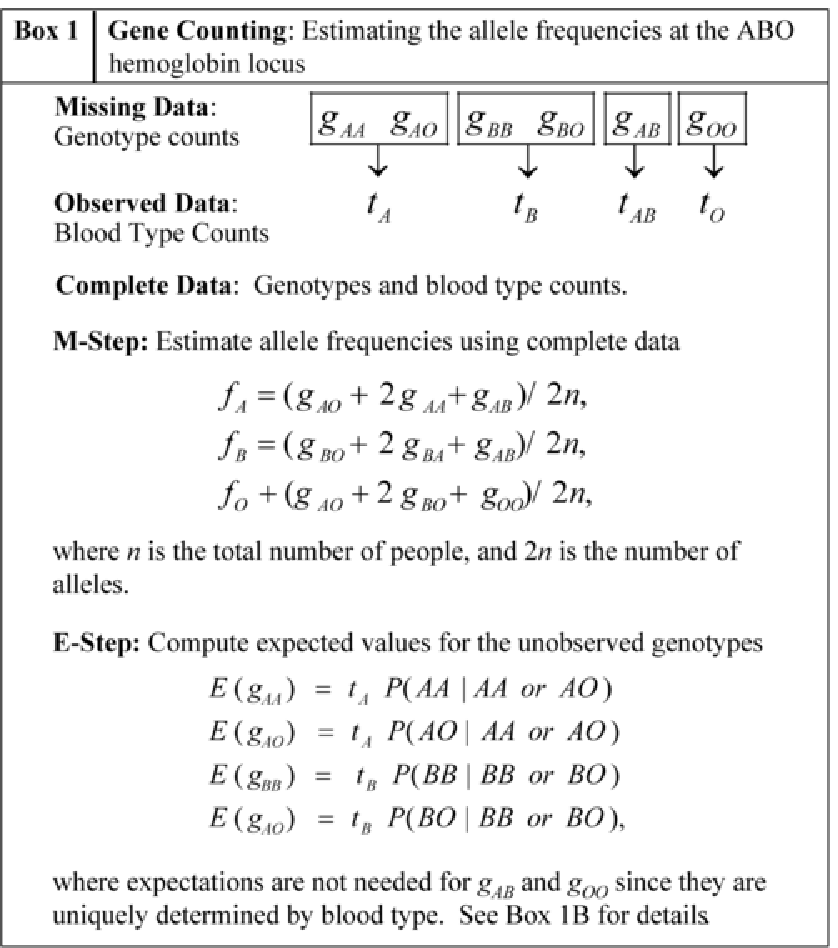}

\end{figure}

\subsection*{Gene Counting}

Box 1 illustrates using the EM algorithm
for estimating the three allele frequencies at the ABO hemoglobin
locus. When the genotype data are directly observed, estimation is
referred to as gene counting, because one simply counts the number
of alleles of each type, and divides by the total number of alleles.
Gene counting with the observed genotype data is shown in the \mbox{M-step}
of Box 1, where the number of individuals (possibly unobserved) with
each genotype is denoted as $g_{AA}$, $g_{AB}$, etc., and the
observed number of individuals with each blood type are denoted as
$t_A$, $t_B$, etc. For an autosomal chromosome, each person
contributes two alleles, hence the denominator of each estimated
frequency is $2n$, where $n$ is the total number of subjects.

The E-step takes into account the known relationships between blood
type and genotype given in the top of Box 1, namely a person with
blood type A must be either AA or AO, and similarly for blood type
B, but blood types AB or O identify genotypes uniquely. For the
E-step, we assume the allele frequencies are known and fixed at the
values estimated at the previous M-step, and use them to calculate
the $P(\mbox{genotype} | \mbox{blood type})$.

The EM algorithm consists of cycling through the two steps,
alternately estimating the allele frequencies assuming the allele
counts are observed, then updating the expected allele counts,
assuming that the frequencies are known. We can start with either
the \mbox{E-~or} the M-step, depending upon whether it is easier to start
with a guess about the frequencies or with a guess about the
genotype counts. Note that in the iterations, the genotypes used at
the M-step are expected, computed at the previous E-step; the allele
frequencies used to compute the conditional expectations at the
E-step are likewise those updated at the previous M-step.

\begin{figure}

\includegraphics{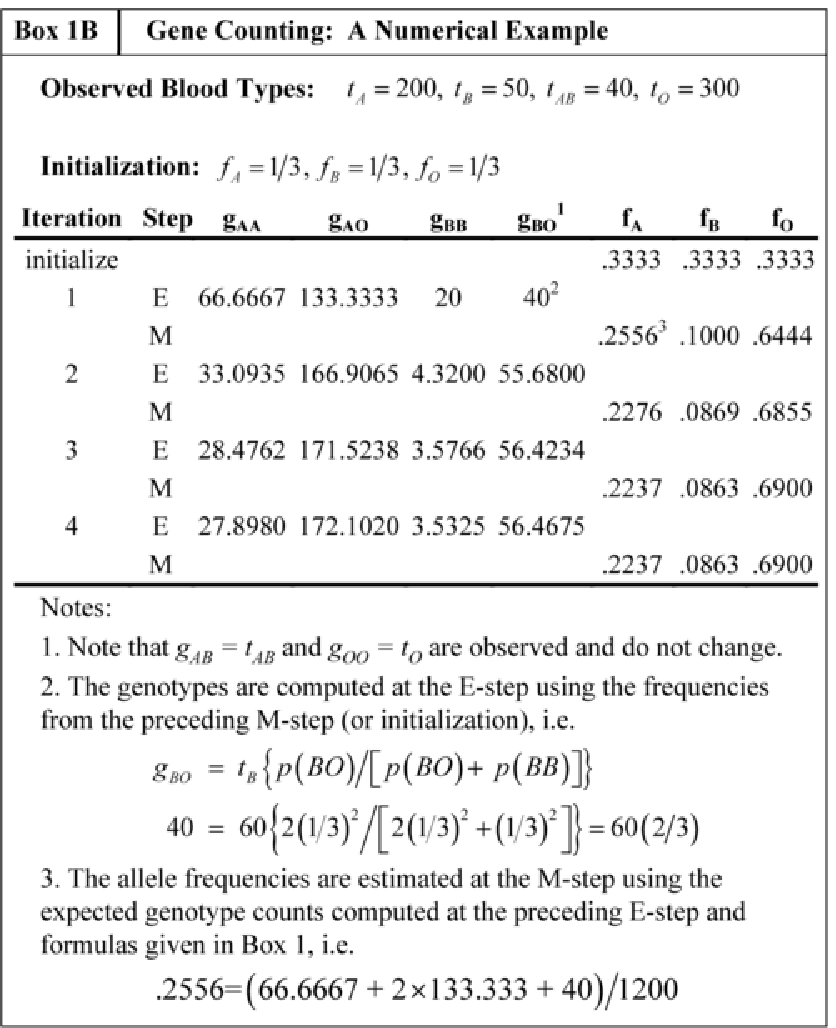}

\end{figure}

Box 1B provides a numerical example to illustrate the algorithm,
using hypothetical, but not unrealistic blood counts from a sample
of 600 subjects. We start the iterations by setting the frequencies
to be $1/3$ each. Computing the expected genotype frequencies at
the E-step uses the assumption of Hardy Weinberg equilibrium to
calculate the probability of genotype frequencies, given allele
frequencies. Hardy Weinberg assumes that allele frequency is the
same for everyone in the population and random mating, hence the two
alleles of an individual are independent. Simple inspection of the
observed data suggests that our initial frequencies are not very
good estimates, but the algorithm converges in only a few
iterations. For actual data examples and further discussion of
using the EM for gene counting, see Lange (\citeyear{lange02}), Chapter 12.

\subsection*{Linkage Example}

Linkage analysis is widely used to find
the chromosomal location of a hypothesized gene affecting some
trait of interest. Simply put, linkage refers to the relative
position of two genetic locations; if they are physically ``close'' on
the same chromosome, the two loci are said to be linked. If they
are on the same chromosome, but distant, or if they are on different
chromosomes, they are said to be unlinked. The EM has many
applications in linkage analysis; here we consider its use in
estimating allele sharing of affected siblings. When two loci are
unlinked, Mendel's laws of inheritance holds independently for the
two loci, and one can easily calculate the probability of allele
sharing between two siblings at a genetic locus. In particular, the
probability that both siblings inherit the exact same alleles from
both their parents (allele sharing is 2) is $\frac{1}{4}$. The
probability that they each inherit two different alleles from both
parents (or allele sharing is zero) is also $\frac{1}{4}$, and the
probability that they inherit the same allele from exactly one
parent (allele sharing is 1) is $\frac{1}{2}$. This type of allele
sharing is called identity-by-descent (IBD) to indicate that the
alleles are shared because the sibs have obtained the same copy from
a common parent.

To implement a linkage analysis, we obtain data on a genetic locus,
called marker data, that we hypothesize is close to a genetic locus
that affects our disease of interest. If both siblings are affected
with the same genetic disease, and the marker and the disease locus
are linked, we expect they are likely sharing at least one allele,
inherited from the same parent, at the unobserved disease locus.
Further, if the locus underlying the disease is linked to the
marker, independent transmission at the 2 loci does not hold.
Rather, we expect that the allele sharing probabilities at the
marker differ from $\frac{1}{4}$, $\frac{1}{2}$, $\frac{1}{4}$, in
the direction of increased sharing. To test this, we estimate the
probabilities of sharing 0, 1 or 2 alleles under $H_A$, $\pi_0$,
$\pi_1$, $\pi_2$ say, from the sample of affected sib pairs and
compare them to the allele sharing probabilities under the null
hypothesis of no linkage, $\frac{1}{4}$,
$\frac{1}{2}$,~$\frac{1}{4}$.

Estimation of the allele sharing frequencies under the alternative
of linkage between the marker and the disease locus is
straightforward if we could observe the IBD sharing directly; we
would simply count the number of affected sibs pairs sharing 0, 1 or
2 alleles, and divide by the number of affected sib pairs. This is
illustrated in the M-step of Box 2 where the complete data IBD
counts are denoted as $Z_0$, $Z_1$ and $Z_2$. As illustrated in Box
2, one cannot always infer IBD sharing from data on parental and
offspring genotypes; it depends upon the pattern of alleles which
are observed. The ``complete data'' illustrates a setting where we can
always deduce IBD sharing, that is, the parents have four distinct
alleles. The ``observed data'' shown in Box 2 illustrates two
situations where we cannot. In general, the ``observed data'' are a
pattern of sharing, for example, either 0 or 1, 1 or 2, etc., together with
the observed parental genotypes. The M-step is based on a
multinomial likelihood, assuming that we observe IBD for each sib
pair, and the E-step provides expectations of the multinomial
counts, by computing the expected allele sharing for each pair
conditional on their observed pattern of sharing, and adding over
pairs. The conditional probability of IBD sharing is obtained from
Bayes rule
\begin{eqnarray*}
&& P  (\mbox{IBD} = j|\mbox{observed data}  )
\\
&&\quad
\propto P  ( \mbox{observed data}|\mbox{IBD}= j  )\pi_j
\quad \mbox{for } j=0,1,2,
\end{eqnarray*}
where $\pi_j$ has been estimated at the M-step; and
$P  ( \mbox{observed data}|\mbox{IBD} = j  )$ is calculated from the observed data
on parents and child's genotypes (Risch, \citeyear{ris90}).

\begin{figure}

\includegraphics{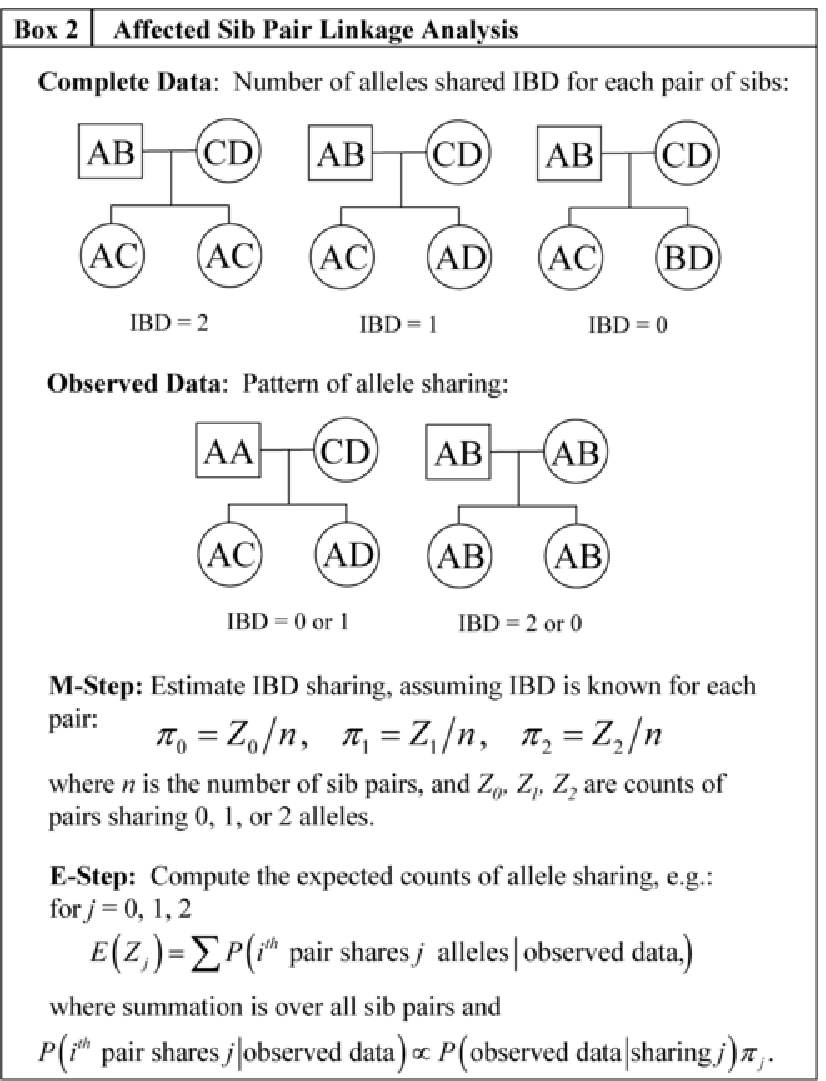}

\end{figure}

Risch (\citeyear{ris90}) also extends the EM to the setting where
parental data are missing, but now one must first have an estimate
of parental allele frequencies.
Kruglyak et al. (\citeyear{kruLan95}) extended the
algorithm to cover multipoint analyses with complex pedigrees as
well as additional markers, but the basic idea of using EM to
estimate IBD probabilities under $H_A$ remains the same.

\section{Example from Computational Biology: Finding Motifs}

Computational Biology deals with the analysis of data that comes
from sequencing the DNA of humans and other organisms. The specific
sequence of the four nucleotides which make up DNA, Adenine,
Guanosine, Thymine and Cytosine, or A, G, C, T, for short, determine
the function of genes and location of genes, the process of RNA
transcription, and the manufacture of proteins essential for cell
function. Understanding many fundamental biological processes
requires tools to identify relatively short patterns of these four
base pairs embedded in long strings of base pairs (approximately 3
billion in the entire human genome). The problems are challenging
and relevant to statisticians because the sequences of interest may
not be ``exact.'' For example, the simple sequence consisting of two
specific base pairs, CG, is relatively rare in the DNA of many
organisms, because CG readily mutates to TG. But mutation is
suppressed in regions near specific genes, forming CG rich islands,
or stretches of DNA that have more CG pairs than ``usual.'' One
approach to identifying GC rich islands is to treat strings of DNA
as realizations of Hidden Markov Models, with different (unobserved)
states corresponding to GC rich or GC poor regions (Parida, \citeyear{par08},
Chapter 5.5; Jones and Pevzner, \citeyear{jonPev04}, Chapter 11). An EM algorithm
to estimate the state and transmission probabilities was developed
by Baum et al. (\citeyear{baumETAL70}).

A related problem in computational biology where EM is used is the
identification of regulatory motifs. Motifs are short sequences of
base pairs, from 6 to 20~pairs in length, which have a similar
pattern of base pairs. Proteins bind to functional motifs located
upstream of genes to encourage the process of RNA transcription in
the genes. Given a set of known genes, the approximate location of
the corresponding functional motifs is known, however their exact
sequences vary because the protein binding process does not require
an exact sequence of base pairs.

The basic idea is illustrated in the top panel of Box~3. Seven
hypothetical DNA fragments are given an input data (Jones and
Pevzner, \citeyear{jonPev04}, Chapter 4).
The underlined portion of each sequence
denotes the actual (unobserved) motif. Here we assume there is only
one motif in each input sequence, and it is known to be exactly
eight base pairs long. The exact DNA letters vary from fragment to
fragment, but two motifs are ``more similar'' than two randomly
selected sequences of eight DNA letters. The problem can be defined
probabilistically by assuming that the probability corresponding to
a letter in a motif location is the same for every motif, but
differs from nonmotif, or ``background'' DNA. The objective is to
characterize the general pattern of DNA for the motifs as a
``consensus'' sequence.

\begin{figure}

\includegraphics{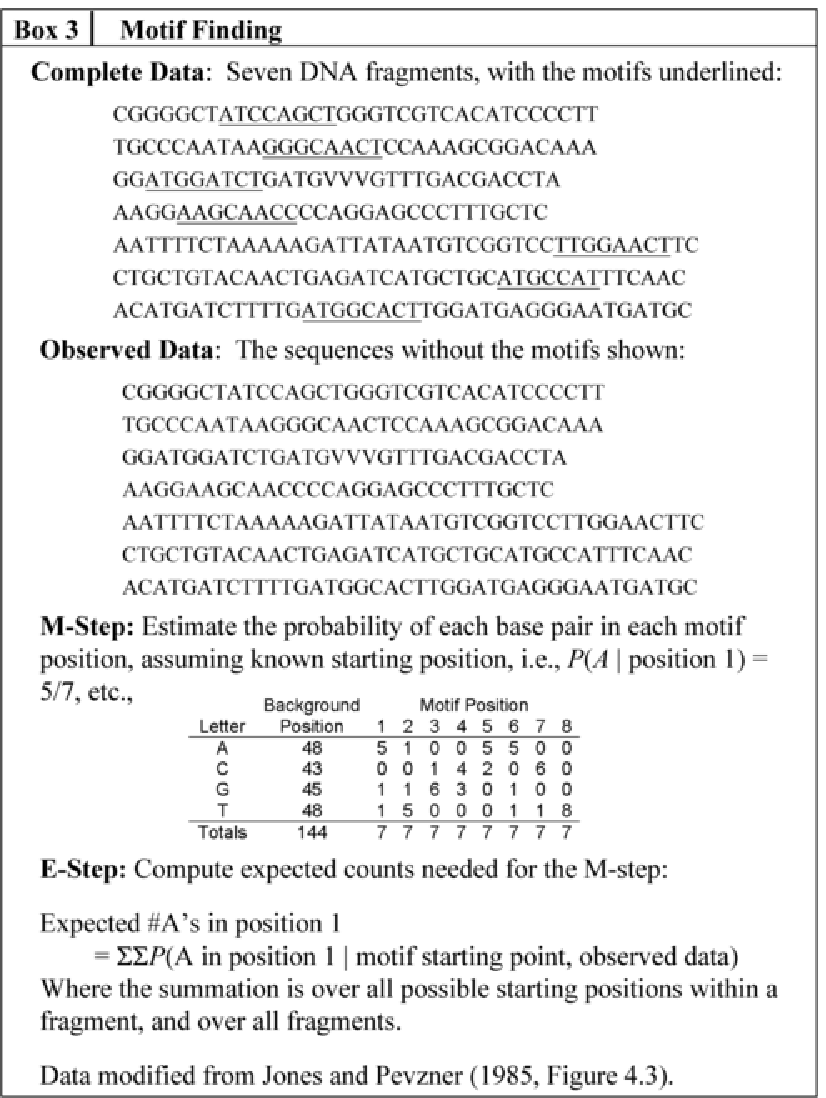}

\end{figure}

A purely computational approach to this problem is to pick a metric
for measuring similarity between eight letter sequences, such as the
number of positions that have the same letter across all fragments,
and then search for the set of eight letter sequences that optimizes
the metric. This is a time consuming search, since each possible
alignment has to be considered, and each fragment of $X$ letters has
$X-7$ possible starting points for the eight letter sequence;
additionally it is difficult to measure optimality of the solution.

Alternatively, one may utilize a probability model for the fragment
data, and use the EM (Lawrence and Reilly, \citeyear{lawRei90}). The data can be
modeled as incomplete because the motif locations are not observed.
Conceptually, the missing data are seven indicator vectors,
indicating the starting point of each motif in each of the seven
fragments. The parameters to be estimated are the frequencies of
the four DNA letters in the motif and nonmotif positions. We
assume a $4\times8$ matrix of multinomial probabilities, one column for
each position in a motif. Each element of the column gives $P(A)$,
$P(C)$, $P(G)$ and $P(T)$ for the DNA letter in each position of the
motif. If we knew the location of each motif in each fragment, we
would estimate these probability vectors via simple multinomial
frequencies for each position of the motif as is illustrated in the
M-step of Box 3.

We do not know the starting points of each motif, but given the
motif and background probabilities, one can readily calculate the
conditional probability of any possible starting position for each
fragment, by assuming a priori that all possible starting values are
equally likely and evaluating the probability of each DNA sequence
for each assumed starting point. These probabilities of each
starting point for each motif are then used to compute the expected
multinomial counts needed for the M-step, as shown in the E-step of
Box 3.

At the conclusion of the iterations of the EM, we have the matrix of
estimated base pair probabilities for each location. These can be
used to compute a ``consensus sequence'' by taking the base pair with
the highest frequency in each location of the motif. For example,
based on the frequencies computed at the M-step in Box 4, the
consensus sequence would be ATGCAACT. At the conclusion of the EM
we also have the probability of each alignment (determined by the
estimated starting points). This has been used to assign motif
locations to specific sequences. While the motif sequence
frequencies can be well estimated with a large number of fragments,
in this formulation there is no \mbox{simplifying} model for alignment
probabilities, and the number of possibilities grows exponentially
with the number of fragments. Hence alignments are unlikely to be
well estimated. This has led to interest in Bayesian approaches
based on Gibbs sampling (Lawrence et al., \citeyear{lawETAL93}). The application of
the EM described here, as in many other settings, is sensitive to
starting values; see Parida (\citeyear{par08}), Chapter 8.6. Generalizations of
the simple case discussed here which allow multiple types of motifs
and as well as multiple numbers per fragment have been given in
Cardon and Stormo (\citeyear{carSto92}) and
Bailey and Elkan (\citeyear{baiElk95}).

\section{Example from Public Health: Monitoring Air Quality}

Many harmful exposures, radon or diesel emissions for example, are
characterized by having particles with very small diameters (less
than 0.4 micrometers). Particles this small cannot be directly
measured, but having estimates of particle sizes are important for
monitoring air quality. Several measurement devices have been
developed to deal with this problem; here we discuss diffusion
batteries which operate on a principle of indirect measurement. The
basic principle is the same used for positron emission tomography
(PET scans) as well as transmission tomography
(Vardi, Shepp and Kaufman, \citeyear{varSheKau85};
Lange and Carson, \citeyear{lanCar84};
Kay, \citeyear{kay97}).

Diffusion batteries are designed to filter out particles of
different sizes by passing a volume of aerosol through a succession
of fine wire mesh screens, and counting the number of particles
remaining in the aerosol at each stage. Figure \ref{f51} illustrates how
the diffusion battery works.

A fixed volume of air is drawn in through the entrance port, with
only one of the exit ports open. The total number of particles
passing through the exit port is counted. Thus the observed data
consists of 11 counts of particles. The ``zero port'' counts the
total number in the volume of aerosol regardless of size, since
there are no screens before the zero port. The subsequent ports
have differing numbers of screens which increase the probability
that particles of different sizes are trapped in the screens, and
are thus not counted. The smaller particles are more likely to be
removed at the early stages, since a particle's diameter determines
how fast they move. The smaller particles are moving faster and are
more likely to hit a barrier (a screen), and become trapped at the
earlier stages. The larger particles are sluggish; they tend to
fall through the battery, only becoming trapped at the end stages of
the battery where there are many more screens. We estimate the
distribution of particle sizes from the total particle counts
measured at each port by dividing the particle size distribution
into intervals, and estimating the proportions in each interval.

\setcounter{figure}{0}
\begin{figure}

\includegraphics{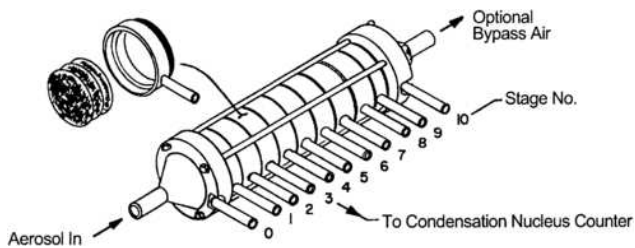}

  \caption{Diagram of a diffusion battery.
Reproduced from the TSI Instruction Manual for
Diffusion Battery Models 3040/3041.\label{f51}}
\end{figure}

A natural way to formulate an incomplete data problem in this
setting is illustrated at the top of Box 4. We define complete data
as a 2-way array of counts of particles in size category $j$ exiting
at the $i$th port, $Z_{ij}$, where $i=1,10$ and
$j=1,8$ in our example. The unobserved $Z_{ij}$ can be modeled as
independent Poisson counts with $E(Z_{ij}) = P_0 w_{ij} f_j$,
where $f_j$ are the frequencies of the $j$th size
category, and
\[
w_{ij} = P(\mbox{particle of size $j$ exits at the $i$th port}).
\]
The $w_{ij}$ are calculated from the known characteristics of the
diffusion battery. The observed counts exiting\vadjust{\goodbreak} each port are just
the row totals of the $Z_{ij}$, $P_i$, plus the count at the zero
port, $P_0$. We note that under this set up, the expected values of
the observed counts follow a simple linear model:
\[
E(P_i) = P_0 \sum_j w_{ij} \times f_j.
\]
Thus estimation can be treated as a linear regression problem where
the $f_j$'s are the coefficients to be estimated and the $w_{ij}$'s
are the known predictors. Because the $f_j$'s are constrained to be
positive, and the errors are not normally distributed, ordinary
least squares does not work well. Typically, non-negative least
squares or Ridge regression have been used as alternatives, but
using the EM to obtain maximum likelihood estimates under the
Poisson model described below represents a substantial improvement
(Maher and Laird, \citeyear{mahLai85}).

\begin{figure}

\includegraphics{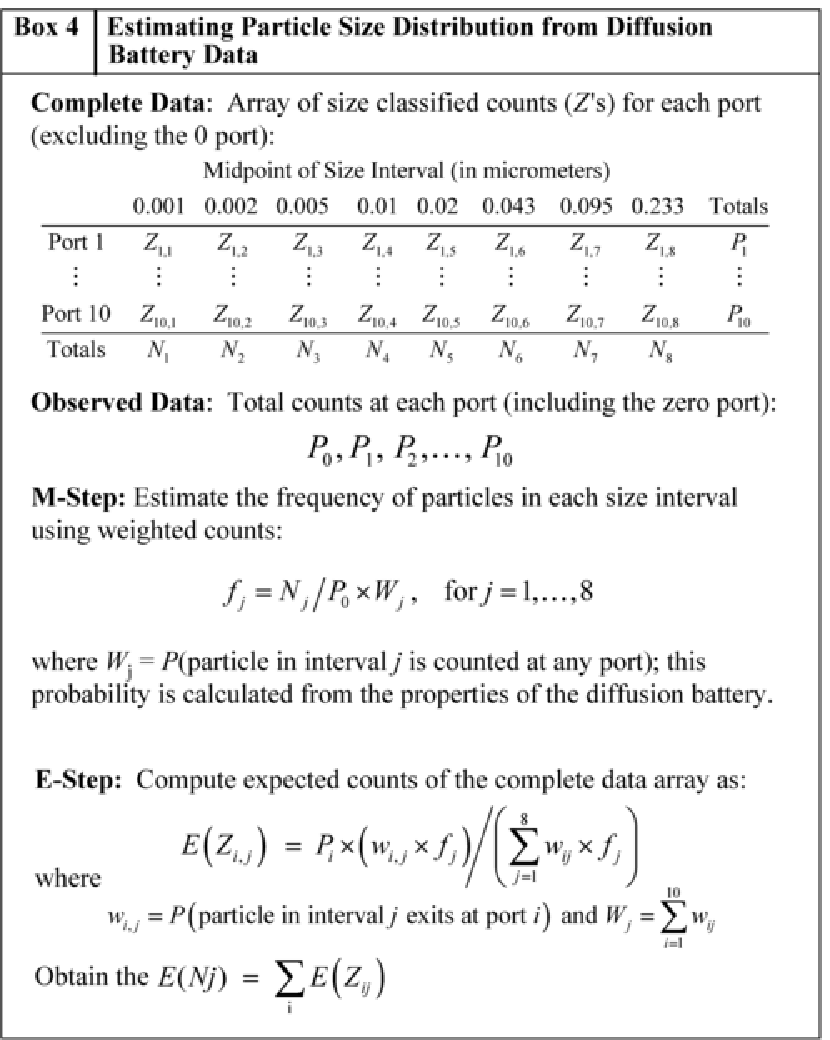}

\end{figure}

The use of the EM for this application is illustrated in Box 4. The
column totals of the array, $N_j$, give the number of particles in a
given size category which exit at all ports combined. Only the
$N_j$'s are needed for the M-step, but taking the full array of
counts as complete data simplifies the calculations. The
proportions, say $f_j$, in each size category are estimated\vadjust{\goodbreak} from the
weighted frequencies $f_j = N_j / P_0 \times W_j$. We use weights,
\begin{eqnarray*}
W_j &=& P(\mbox{particle of size } j \mbox{ is counted at any port})
\\
&=& \sum^{10}_{i=1} w_{ij},
\end{eqnarray*}
because the complete data counts have been filtered
according to size.

At the E-step, we compute the expected values of each $Z_{ij}$,
conditioning on the observed row margins and assuming the $f_j$'s
are known.

\subsection*{Conclusion}

This paper merely skims the surface of the
multitude of applications in diverse scientific areas where the EM
plays an important role, not just in the computations, but in the
conceptualization of the problem as well. This volume provides
additional examples from other areas of the empirical sciences.

\vspace*{-1pt}
\end{document}